# Resilience based optimization for western U.S. transmission grid against cascading failures

**Jian Zhou, Stamatis Tsianikas and David W. Coit**
**Department of Industrial & Systems Engineering, Rutgers University**
**Piscataway, NJ 08854, USA**

**Frank A. Felder**
**Edward J. Bloustein School of Planning and Public Policy, Rutgers University**
**New Brunswick, NJ 08901, USA**

## Abstract

Real-world network systems, for example, power grids, are critical to modern economies. Due to the increasing system scale and complex dependencies inside of these networks, system failures can widely spread and cause severe damage. We have experienced massive cascading failures in power grids, such as major U.S. western grid failures in 1996 and the great Northeast blackout of 2003. Therefore, analyzing cascading failures and defense strategies, in response to system catastrophic breakdown, is crucial. Although many efforts have been performed to prevent failure propagation throughout systems, optimal system restoration considering system dependency against cascading failures is rarely studied. In this paper, we present a framework to optimize restoration strategies to improve system resiliency regarding cascading failures. The effects of restoration strategies are evaluated by system resilience loss during the cascading process. Furthermore, how system dependency influences the effects of the system restoration actions against cascading failures are investigated. By performing a case study on the western U.S. transmission grid, we demonstrate that our framework of system restoration optimization can enhance system resiliency by reducing the intensity and extent of cascading failures. Our proposed framework provides insights regarding optimal system restoration from cascading failures to enhance resiliency of real-life network systems.

## 1. Introduction

Large-scale network systems, such as electricity power systems, communication networks and traffic networks, are becoming increasingly prevalent and interconnected [1, 2]. These network systems are playing significant roles in people's daily life and industrial development. As a result, ensuring that these systems work effectively and reliably is of great concern [3, 4]. Although many efforts have been made to prevent damage of these interdependent systems, the previous decades have witnessed world-wide various high-impact failures on different network systems following disruptive events, especially in power grids [5-7]. For example, the extensive power outages caused by Superstorm Sandy in October 2012 affected more than 8.2 million people across 21 states in the U.S. East Coast in terms of economic and social destruction [8]. Consequently, developing effective system restoration strategies to enhance power grid resiliency and avoid further failure propagation becomes an important and exciting research topic [9-11].

Sun et al. [12] developed a new reliability model for networks with fractal growth and no-loop (NF-NL). The model is based on the fractal unit of a system, which has a bigger scale than any individual system node or edge, and a much smaller scale than the whole network. Zhou et al. [13] proposed a simulated based method for optimizing battery size added to photovoltaic (PV) array systems to make power system grid-outage resilient and economically viable. Previous system interruptions have already shown the remarkable influence of multiple system dependencies on system robustness and dynamic failure behaviors [14, 15]. However, current research on system resiliency enhancement rarely if ever takes into account the relationship between multiple system dependencies and different restoration strategies regarding cascading failures.

In this paper, two different restoration strategies are applied to western states U.S. power transmission grid, which is subject to the mixed cascading failures. The impacts of two system dependence characteristics, i.e., the distribution that dependence cluster sizes follow and dependency strength, combined with restoration strength, on system resiliency and restoration effects are investigated. Then, optimal restoration strategies can be selected based on the obtained results.

## 2. Restoration strategies

In this paper, system dependence clusters and system dependency strength are used to describe system dependency. System dependence cluster is defined as a collection of system components/nodes which have dependency relations between each other, for instance, functional dependency [16]. The illustration of single network system that contains dependence clusters of components/nodes is shown in Fig. 1. Dependence cluster collapsing threshold (*CCT*) denotes the dependency strength. It means that a dependence cluster of network nodes instantly collapses, i.e., all nodes belonging to this dependence cluster fail once the proportion of failed nodes of this cluster exceeds *CCT* [17]. Therefore, larger *CCT* indicates weaker dependency relationship among network nodes. It means that the node failure in a dependence cluster has less impact on other functioning nodes belonging to the same dependence cluster.

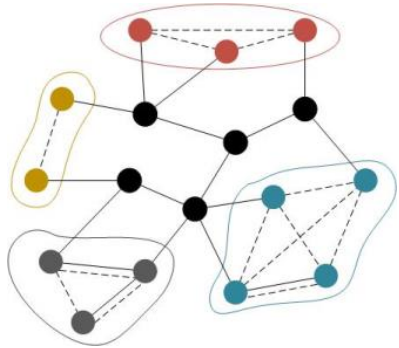

Fig. 1. Network system with dependence clusters of nodes

A cascading failure model, which was designed for single networks subject to mixed cascading failures including dependence-caused failures and load dynamics-caused failures [16], is used to describe cascading failure process. Restoration strategies are applied to network systems after cascading failures occur. In this paper, the main procedures for network system restoration are implemented and evaluated on the basis of the optimal restoration model [18].

Two restoration strategies are applied in this paper, same repair probability strategy (SRP) and High-degree first repair strategy (HDFR). SRP restoration strategy means the failed network nodes, where repair activity has not yet started, are randomly selected to be repaired when maintenance resources are available to do so. The proportion of failed nodes which are selected to start maintenance when resources are available is decided by $R_p$. It is a parameter that indicates the restoration strength and the capability of assigned or available maintenance resources. Larger $R_p$ means stronger restoration strength. SRP restoration strategy is considered as a baseline to be compared with other restoration strategies. HDFR restoration strategy means that the failed nodes of which repairing has not yet started are repaired in a descending order of node degree, i.e., network nodes with more node degrees are repaired with higher priority.

The metrics adopted to evaluate restoration effects and system resiliency are *StpLCC*, *StpPerLd* and *SLF*. *StpLCC* is the size of the largest connected component (*LCC*) of the network system when the system is recovered to the predetermined level. *LCC* is the largest group of network nodes which are fully connected with each other. Larger *StpLCC* indicates better network connectivity. *StpPerLd* is the percentage of total network load when the network is recovered to the predetermined level to the initial amount of network load. This metric reflects the recovery of network transmission capability, that is, larger *StpPerLd* indicates better transmission capability. *SLF* represents total network load fluctuations including cascading failures and restoration process until network performance is restored to the acceptable level. It is calculated by Eq. (1),

$$SLF = \sum_{i=1}^{T} |TL(t=i) - TL(t=0)|, TL(t=i) = \sum_{j=1}^{N} L_j(t=i) \quad (1)$$

where $L_j(t = i)$ denotes the load on node $j$ at $t = i$, and $L_j(t = i) = 0$ if node $j$ is failed at $t = i$. After the failure of a node the load is instantaneously reorganized on the rest of the network system. $TL(t = i)$ means total network load at $t = i$. $TL(t = 0)$ is total network load at initial time, $t = 0$, before any failures occur. As a result, *SLF* measures system resilience loss in terms of network load when system performance is recovered to a predetermined level at time *T*, which is a type of system performance taking into account system load demand and supply capability. Larger *SLF*, i.e., larger system resilience loss, indicates lower system resiliency.

### 3. Simulation analysis of the western U.S. transmission grid

In this paper, we perform simulation of cascading failures and restoration actions to an infrastructural system, the western U.S. power transmission grid [19], which has already been used by many researchers [20]. This realistic power system has a topology with 4,941 nodes and 6,594 edges. Each node represents a power plant, a transformer or a consumer. Two nodes are connected by undirected edges, on which the electric current can flow in both directions, if they are physically connected by cables. The average edges of each node (node degree) in this network system are 2.67 and most nodes have comparable degrees. The capacity of each network node in the western U.S. power transmission grid is assigned according to the node capacity-load model [21]. In this paper, the two model parameters for node capacity allocation, the tolerance parameter of networks, $\alpha$, is equal to 1.1, and the nonlinear coefficient, $\beta$, is equal to 0.5.

In this section, we analyzed the impacts of restoration strength, $R_p$, and two dependence characteristics, i.e., *CCT* and the distribution of sizes of dependence clusters, on the effects of two restoration strategies and system resiliency regarding mixed cascading failures.

Two different assumptions of dependence cluster distributions are considered in this paper, respectively, (1) each dependence cluster in the system has the same number of dependent nodes, which is *D-size*; (2) sizes of dependence clusters follow an adjusted Poisson distribution (to prevent clusters with zero components/nodes), where the mean size is *D-size* [22].

For the following results, *D-size* = {3, 5, 7}, *CCT* = {0.5, 0.9}, $R_p$ = {0.7, 0.8}. The required repair time of each failed component/network node, $T_r = 1$, is assumed to be constant. Random failures of network nodes, that is, initial random failed network nodes, are triggers for cascading failures. Simulation results are averaged over different number of initial failed system nodes, and the initial failed numbers of system nodes that we considered are {9, 15, 21, 27, 33, 39, 45, 51, 57, 63}. Results are presented in the following Tables 1 and 2. Note that each value in the tables corresponds to the average over 5 simulation realizations. It is assumed that the entire simulation process, including cascading failures and restoration, stops when network efficiency is recovered to be more than 90% of the initial system state.

#### 1) Results under SRP strategy
SRP restoration strategy is applied first. The corresponding results of three metrics are shown in Table 1a to c.

Table 1a: Average SRP results with dependence cluster size following different distributions - *StpLCC*

| *StpLCC* | All clusters with same fixed size | | | | | Cluster sizes follow adjusted Poisson distribution | | | | |
|---|---|---|---|---|---|---|---|---|---|---|
| | $R_p$=0.7 | | $R_p$=0.8 | | Average | $R_p$=0.7 | | $R_p$=0.8 | | Average |
| *CCT* | 0.5 | 0.9 | 0.5 | 0.9 | | 0.5 | 0.9 | 0.5 | 0.9 | |
| *D-size*=3 | 4761 | 4809 | 4755 | 4770 | 4774 | 4713 | 4745 | 4867 | 4824 | 4787 |
| *D-size*=5 | 4816 | 4685 | 4736 | 4822 | 4765 | 4695 | 4795 | 4806 | 4762 | 4765 |
| *D-size*=7 | 4622 | 4797 | 4826 | 4797 | 4761 | 4692 | 4726 | 4844 | 4720 | 4746 |
| Average | 4733 | 4764 | 4772 | 4796 | 4766 | 4700 | 4755 | 4839 | 4769 | 4766 |

Table 1b: Average SRP results with dependence cluster size following different distributions - *StpPerLd*

| StpPerLd (%) | All clusters with same fixed size | | | | | Cluster sizes follow adjusted Poisson distribution | | | | |
|---|---|---|---|---|---|---|---|---|---|---|
| | $R_p$ =0.7 | | $R_p$ =0.8 | | Average | $R_p$ =0.7 | | $R_p$ =0.8 | | Average |
| CCT | 0.5 | 0.9 | 0.5 | 0.9 | | 0.5 | 0.9 | 0.5 | 0.9 | |
| D-size=3 | 97.12 | 98 | 97.06 | 96.62 | 97.2 | 96.02 | 96.34 | 98.58 | 97.4 | 97.1 |
| D-size=5 | 97.44 | 95.92 | 96.22 | 97.62 | 96.8 | 95.8 | 97.64 | 96.9 | 95.78 | 96.5 |
| D-size=7 | 95 | 97.32 | 97.36 | 97.42 | 96.8 | 95.1 | 95.4 | 98.4 | 96.5 | 96.4 |
| Average | 96.5 | 97.1 | 96.9 | 97.2 | 96.9 | 95.6 | 96.5 | 98.0 | 96.6 | 96.7 |

Table 1c: Average SRP results with dependence cluster size following different distributions - *SLF*

| SLF (×10³) | All clusters with same fixed size | | | | | Cluster sizes follow adjusted Poisson distribution | | | | |
|---|---|---|---|---|---|---|---|---|---|---|
| | $R_p$ =0.7 | | $R_p$ =0.8 | | Average | $R_p$ =0.7 | | $R_p$ =0.8 | | Average |
| CCT | 0.5 | 0.9 | 0.5 | 0.9 | | 0.5 | 0.9 | 0.5 | 0.9 | |
| D-size=3 | 12502 | 8946 | 13398 | 14862 | 12427.0 | 17434 | 16200 | 6097 | 11285 | 12754.0 |
| D-size=5 | 11264 | 17952 | 16734 | 10549 | 14124.8 | 18434 | 10532 | 13606 | 18471 | 15260.8 |
| D-size=7 | 22042 | 11872 | 11586 | 11578 | 14269.5 | 21790 | 20012 | 6938 | 15294 | 16008.5 |
| Average | 15269.3 | 12923.3 | 13906.0 | 12329.7 | 13607.1 | 19219.3 | 15581.3 | 8880.3 | 15016.7 | 14674.4 |

As indicated in Table 1, dependence cluster size following different distributions does not noticeably impact the effect of SRP restoration strategy in terms of *StpLCC* and *StpPerLd* except resilience loss, i.e., *SLF*. In comparison to resilience loss accompanied with system having dependence clusters with same fixed size, greater resilience loss is incurred when dependence cluster sizes follow an adjusted Poisson distribution.

In addition, larger $R_p$ or smaller *D-size* lead to larger *StpLCC*, *StpPerLd* and smaller *SLF*, which indicates better system connectivity, transmission capability and less resilience loss denoting better restoration effects. Whereas the impacts of *CCT* on restoration effects are distinct when dependence cluster sizes follow different assumptions. Larger *CCT* contributes to better restoration effects when all dependence clusters have same fixed size, while *CCT* impact is influenced by $R_p$ when cluster sizes follow an adjusted Poisson distribution.

### 2) Results under HDFR restoration strategy

We applied HDFR restoration strategy, and the results of three metrics in this case are shown in Table 2a to c.

Table 2a: Average HDFR results under dependence cluster size following different distributions - *StpLCC*

| StpLCC | All clusters with same fixed size | | | | | Cluster sizes follow adjusted Poisson distribution | | | | |
|---|---|---|---|---|---|---|---|---|---|---|
| | $R_p$ =0.7 | | $R_p$ =0.8 | | Average | $R_p$ =0.7 | | $R_p$ =0.8 | | Average |
| CCT | 0.5 | 0.9 | 0.5 | 0.9 | | 0.5 | 0.9 | 0.5 | 0.9 | |
| D-size=3 | 4920 | 4910 | 4932 | 4924 | 4922 | 4904 | 4910 | 4917 | 4927 | 4915 |
| D-size=5 | 4911 | 4914 | 4927 | 4932 | 4921 | 4912 | 4914 | 4928 | 4931 | 4921 |
| D-size=7 | 4913 | 4921 | 4931 | 4929 | 4924 | 4906 | 4916 | 4929 | 4924 | 4919 |
| Average | 4915 | 4915 | 4930 | 4928 | 4922 | 4907 | 4913 | 4925 | 4927 | 4918 |

Table 2b: Average HDFR results under dependence cluster size following different distributions - *StpPerLd*

| StpPerLd (%) | All clusters with same fixed size | | | | | Cluster sizes follow adjusted Poisson distribution | | | | |
|---|---|---|---|---|---|---|---|---|---|---|
| | $R_p$=0.7 | | $R_p$=0.8 | | Average | $R_p$=0.7 | | $R_p$=0.8 | | Average |
| CCT | 0.5 | 0.9 | 0.5 | 0.9 | | 0.5 | 0.9 | 0.5 | 0.9 | |
| D-size=3 | 99.46 | 99.24 | 99.78 | 99.72 | 99.6 | 98.74 | 99.32 | 99.06 | 99.76 | 99.2 |
| D-size=5 | 99.2 | 99.32 | 99.68 | 99.82 | 99.5 | 99.38 | 99.44 | 99.66 | 99.84 | 99.6 |

| *D-size*=7 | 99.32 | 99.48 | 99.72 | 99.76 | 99.6 | 98.76 | 99.34 | 99.74 | 99.78 | 99.4 |
| Average | 99.3 | 99.3 | 99.7 | 99.8 | 99.5 | 99.0 | 99.4 | 99.5 | 99.8 | 99.4 |

Table 2c: Average HDFR results under dependence cluster size following different distributions - *StpLCC*

| *SLF* ($\times 10^3$) | All clusters with same fixed size | | | | | Cluster sizes follow adjusted Poisson distribution | | | | |
|---|---|---|---|---|---|---|---|---|---|---|
| | *Rp* =0.7 | | *Rp* =0.8 | | Average | *Rp* =0.7 | | *Rp* =0.8 | | Average |
| *CCT* | 0.5 | 0.9 | 0.5 | 0.9 | | 0.5 | 0.9 | 0.5 | 0.9 | |
| *D-size*=3 | 2393 | 3331 | 1424 | 1908 | 2264.0 | 5625 | 3332 | 4528 | 1714 | 3799.8 |
| *D-size*=5 | 3637 | 3253 | 1663 | 1399 | 2488.0 | 2965 | 2738 | 2136 | 1431 | 2317.5 |
| *D-size*=7 | 2934 | 2617 | 1595 | 1651 | 2199.3 | 5606 | 2931 | 1695 | 1860 | 3023.0 |
| Average | 2988.0 | 3067.0 | 1560.7 | 1652.7 | 2317.1 | 4732.0 | 3000.3 | 2786.3 | 1668.3 | 3046.8 |

Similar to what was found from Table 1, the distribution that dependence cluster sizes follow does not markedly influence the restoration effect of HDFR strategy regarding *StpPerLd* and *StpLCC*. However, the distribution of cluster sizes affects resilience loss, i.e., *SLF*. Less resilience loss is achieved for a system that has dependence clusters with the same fixed size compared with that of the system with dependence cluster sizes following an adjusted Poisson distribution.

It can be seen from Table 2 that larger $R_p$ leads to better restoration effect in terms of three metrics without respect to *D-size* or *CCT* is, whereas *D-size* does not show a monotonic impact on restoration effect. Larger *CCT* contributes to better restoration effect when dependence cluster sizes follow an adjusted Poisson distribution. While *CCT* has an opposite impact on restoration effect when dependence clusters have same fixed size, which means larger *CCT* impairs restoration effect.

Above all, it can be observed that HDFR restoration strategy shows better effects regarding resiliency recovery than that of SRP restoration strategy on western states U.S. power transmission grid against mixed cascading failures.

## 4. Conclusions

In this paper, western states U.S. power transmission grid is used as the network system to conduct analysis of cascading failures and restoration process. According to the results from the application of two restoration strategies, restoration strength, which is presented by $R_p$, exerts a remarkable impact on restoration effect with respect to network system connectivity, transmission capability and system resilience loss. Larger $R_p$ leads to better restoration effect, while the impacts of network dependence characteristics on restoration effects are distinct when applying different restoration strategies. It demonstrates that it is critical to taken into account the specific dependence characteristics with the network systems to find optimal restoration strategy. Our results can be useful for understanding and selecting optimal restoration strategy for different network systems. In the future research, other restoration strategies can be applied to different real-world network systems considering the impacts of multiple network dependencies. Since available budgets for restoration investment are limited in real cases, budget constraints can be incorporated regarding restoration optimization from an economic aspect in the next extension of current research.